\documentclass[aps,prc,showpacs,showkeys,twocolumn]{revtex4-1}
\usepackage{epsfig}
\usepackage{amssymb}
\usepackage{amsmath}
\usepackage{subeqnarray}
\usepackage{graphicx}
\newcommand{\be}{\begin{equation}}
\newcommand{\ee}{\end{equation}}
\newcommand{\ben}{\begin{eqnarray}}
\newcommand{\een}{\end{eqnarray}}

\usepackage{color}

\begin{document}

\title{\bf Critical behaviour of an effective relativistic mean field model in the presence of magnetic background and boundaries}
\author{L. M. Abreu$^{1}${\footnote{email: luciano.abreu@ufba.br}},
E. S. Nery$^{1}${\footnote{email: elenilsonnery@hotmail.com}}
}
\affiliation{$^{1}$Instituto de F{\'i}sica, Universidade Federal da
Bahia, 40210-340, Salvador-BA, Brazil}

\begin{abstract}
In the present work we investigate the combined influence of magnetic background and boundaries on the thermodynamic properties of effective relativistic mean field models, like the so-called Walecka model. This is done by making use of generalized zeta-function approach and mean-field approximation at effective chemical equilibrium, focusing on the dependence with the size of compactified spatial dimension, the temperature and the magnetic field strength. The findings suggest a rich phase structure in the parameter space. The maintenance of long-range correlations is strongly affected under the change of these parameters, with the symmetric phase being favoured due to both inverse magnetic catalysis effect and the reduction of size of compactified dimension.

\end{abstract}

\maketitle
%%%%%%%%%%%%%%%%%%%%%%%%%%%%%%%%%%%%%%%%%%%%%%%%%%%%%%%%%%%%%%%%%%%%%%%%%%%%%%%%%%%%%%%%
%%%%%%%%%%%%%%%%%%%%%%%%%%%%%%%%%%%%%%%%%%%%%%%%%%%%%%%%%%%%%%%%%%%%%%%%%%%%%%%%%%%%%%%%
\section{INTRODUCTION}
%%%%%%%%%%%%%%%%%%%%%%%%%%%%%%%%%%%%%%%%%%%%%%%%%%%%%%%%%%%%%%%%%%%%%%%%%%%%%%%%%%%%%%%%
%%%%%%%%%%%%%%%%%%%%%%%%%%%%%%%%%%%%%%%%%%%%%%%%%%%%%%%%%%%%%%%%%%%%%%%%%%%%%%%%%%%%%%%%

Experimental and theoretical advances over the past few decades have afforded us a unique opportunity to assess strongly interacting matter under extreme conditions. The understanding of its thermodynamic and transport properties remains at the vanguard of our efforts destined to depict its rich phase structure at finite temperature and  baryon number density~\cite{Bellac,Kapusta,Ashok}.

In the theoretical scenario, effective quantum field theories of Quantum Chromodynamics (QCD) at finite temperature have been very useful tools in the mapping of strongly interacting matter. 
In particular, an illustrative case is the Walecka-like model~\cite{Walecka:1974qa}. One of its qualities is the (at least qualitative) simple description of a large number of phenomena.
Its prototypical system is the nuclear matter with exchange nucleon-nucleon interactions. In this context of finite temperature field theory, it means a gas of nucleons immersed in a hot and dense hadronic medium  constituted of light mesons and other particles which induce attractive and repulsive short-range interactions~\cite{JTheis:1983PRD,Saito,Menezes0,Delfino,Lavagno,Shao,Casali,Fukushima,Dutra,Torres2,Zhang:2017etr,Oertel:2016bki}. In this sense, thermodynamic behaviour of this system can be investigated when it is under certain conditions, like finite temperature, finite chemical potential, and other. The major part of the theoretical studies assumes an infinite volume without magnetic fields.

Looking specially at the influence of external magnetic field on the phase diagram, the motivation comes from the phenomenology of heavy ion collisions and compact stars, in which a strong magnetic background is produced~\cite{Kharzeev,Skokov:2009qp,Chernodub:2010qx,Ayala1,Tobias,Heber,MAO,Ayala2,Mamo:2015dea,Pagura,Magdy,Ayala0}. In the case of heavy-ion collisions at RHIC and LHC, for example, the analyses suggest a strong dependence of phase diagram with the strength of magnetic field, which is estimated to be of the hadronic scale, that is of the order of $1-10\; m_{\pi}^2$ or even higher ($ m_{\pi} = 140$ MeV is the pion mass) and with duration expected to be rather short (a few fm/c). 

 Besides, the importance of finite-volume effects on the phase diagram of strongly interacting matter has also been a subject of great interest. A large amount of effort has been devoted to estimate how the long-range correlations are  affected due to finite-size effects and their implications on the thermodynamic quantities and transport properties  in various contexts~\cite{Kim,Braun,Abreu:2006,Ebert0,Abreu:2009zz,Boomsma,Skokov,Gatto,Ebert1,Palhares2,Abreu:2011rj,Braun3,Palhares,Luciano1,Braun2,Flachi,Ebert2,Fraga,Bhattacharyya1,Abreu3,Abreu6,Ebert3,Bhattacharyya2,Bhattacharyya3,Abreu4,Abreu5,Abreu7,Bao1,PhysRevC.96.055204,Samanta,Wu,Shi}. Notice, particularly, that in Ref.~\cite{PhysRevC.96.055204} is performed an investigation about the relevance of the boundaries on  Walecka's mean-field theory without magnetic field, and the findings suggest that the thermodynamic behaviour depends on the length and number $d$ of compactified spatial dimensions, with the symmetric phase being favoured as the size of the system diminishes and $d$ increases. In other words,  it has been shown there that the critical temperature for the crossover transition at effective chemical equilibrium decreases as the volume decreases.
 
 It is worthy mentioning that the influence of a finite-volume and the presence of a strong magnetic field has been discussed in the literature in different effective approaches~\cite{Fraga,Boomsma,Bhattacharyya1,Abreu3,Bhattacharyya3,Skokov,Gatto,Shi}. Accordingly, a natural question arises about the phase structure of a  thermal gas of fermions interacting with a medium constituted of light hadrons  (described as a first approximation by Walecka model), confined in a reservoir and under the effect of a magnetic background. 

Thus, in the present work we are interested in the combined influence of finite size and magnetic background on the thermodynamic properties of a fermionic system, modelled as a first approximation by the Walecka's mean-field theory at finite temperature. We employ the zeta-function regularization approach to take into account the finite temperature and size of compactified spatial dimension, and obtain the thermodynamic potential and solutions of gap equations. The phase diagram is analyzed under the change of relevant parameters. 

The paper is organized as follows. Section II, we present the Lagrangian density and obtain the thermodynamical quantities of Walecka model in mean-field approximation via the zeta function regularization approach. A discussion of the results and phase structure is done in Section III. Finally, Section IV is devoted to concluding remarks.

%%%%%%%%%%%%%%%%%%%%%%%%%%%%%%%%%%%%%%%%%%%%%%%%%%%%%%%%%%%%%%%%%%%%%%%%%%%%%%%%%%%%%%%%
%%%%%%%%%%%%%%%%%%%%%%%%%%%%%%%%%%%%%%%%%%%%%%%%%%%%%%%%%%%%%%%%%%%%%%%%%%%%%%%%%%%%%%%%
\section{THE FORMALISM}
%%%%%%%%%%%%%%%%%%%%%%%%%%%%%%%%%%%%%%%%%%%%%%%%%%%%%%%%%%%%%%%%%%%%%%%%%%%%%%%%%%%%%%%%
%%%%%%%%%%%%%%%%%%%%%%%%%%%%%%%%%%%%%%%%%%%%%%%%%%%%%%%%%%%%%%%%%%%%%%%%%%%%%%%%%%%%%%%%
We start by introducing the Walecka model in the presence of an external magnetic field, whose Lagrangian density is given by  
\begin{eqnarray}
\mathcal{L}&=&\bar{\psi} \left(i\gamma^{\mu}\tilde{D}_{\mu} -m_{\psi}+g_{\sigma}\sigma \right)\psi+\nonumber\\
&&+\frac{1}{2} \left(\partial_{\mu}\sigma\partial^{\mu}\sigma-m_{\sigma}^2\sigma^2\right) - \frac{1}{4}W^{\mu\nu}W_{\mu\nu}+\nonumber\\
&&+\frac{1}{2}m_{\omega}^2\omega_{\mu}\omega^{\mu},\label{eqII1}
\end{eqnarray}
where $\tilde{D}_{\mu} = \partial_{\mu} + i g_{\omega}\gamma^{\mu}\omega_{\mu} +ieA^{ext}_{\mu}$ is the modified covariant derivative, with $A^{ext}_{\mu}$ denoting the four-potential associated to the magnetic background; $m_{\psi}$, $m_{\sigma}$ and $m_{\omega}$ are the masses of fermion, scalar and vector fields, respectively;  $W_{\mu\nu}=\partial_{\mu}\omega_{\nu}-\partial_{\nu}\omega_{\mu}$ is the $\omega$-field strength tensor; and $g_{\sigma }$  ($g_{\omega }$) is the coupling constant for interaction between and Dirac and scalar (vector) field.

The equations of motion can be obtained from Eq.~(\ref{eqII1}); in Lorentz gauge ($\partial_{\mu}\omega^{\mu}=0$) they have the following form,
\begin{eqnarray}
\left[ i\gamma^{\mu}\tilde{D}_{\mu}-(m_{\psi}^2-g_{\sigma}\langle\sigma\rangle) \right]\psi & = & 0, \label{eqII2} 
\\
(\partial_{\mu}\partial^{\mu}+m_{\sigma}^2)\sigma & = & g_{\sigma}\rho_s,\label{eqII3} \\
(\partial_{\nu}\partial^{\nu}+m_{\omega}^2)\omega_{\mu} & = & ig_{\omega}j_{\mu}.\label{eqII4}
\end{eqnarray}
In Eqs.~(\ref{eqII3}) and (\ref{eqII4}), $\rho_s=\bar{\psi}\psi$ denotes the scalar density, while $j^{\mu}=\bar{\psi}\gamma ^{\mu}\psi$ is the fermion 4-current.

In the present analysis we are interested in the lowest-order estimate of the thermodynamic properties of the $\psi$-field. Then, its interactions with other fields will be performed within the mean-field approximation, which means that we will neglect the fluctuations of the scalar and vector 
fields. The $\sigma$ and $\omega$ fields are replaced by 
their classical counterparts: $\sigma = \langle\sigma\rangle$ and $\omega =\langle\omega^0\rangle$, with $\omega^{\mu}=0$ for $\mu\neq0$.
Therefore, after some manipulations the equation
of motion for the $\psi$-field minimally coupled to the external magnetic field [Eq.~(\ref{eqII2})] can be rewritten as
\begin{eqnarray}
[\vec{D}^2-e(\vec{\sigma}\cdot\vec{B})+(i\partial_0-g_{\omega}\langle\omega_0\rangle)^2-M_{*}^2]\psi=0,\label{eqII5}
\end{eqnarray}
where $\vec{D}=\vec{p}-e\vec{A}$, $\vec{\sigma}$ are the Pauli matrices, and \begin{eqnarray}
M_{*}=m_{\psi}-g_{\sigma}\langle\sigma\rangle\label{massa}
\end{eqnarray}
is the effective mass.

For convenience, we choose for the magnetic background the Landau gauge $A^{\mu}_{ext}=(0,-Hx_2,0,0)$, where $H$ is the intensity of 
the uniform external magnetic field in the direction $z$. In this context, the field operators are given by the set of the normalized eigenfunctions of the Landau basis. This means that the solutions of equation of motion in Eq.~(\ref{eqII5}) read,
\begin{eqnarray}
\psi(x)=e^{i(p_0x_0-p_1x_1-p_3x_3)}u(x_2),\label{eqII6}
\end{eqnarray}
where $u(x_2)$ satisfies a harmonic oscillator equation,
\begin{eqnarray} 
\left[-\partial_{x_2}^2+e^2H^2\left(x_2-\frac{p_1}{eH}\right)^2\right]u(x_2)=0,\label{eqII7}
\end{eqnarray}
whose solutions are given by
\begin{eqnarray}
u(x_2)=\frac{1}{\sqrt{2^{l}l!}}\left(\frac{eH}{\pi}\right)^{1/4}H_{l}\left(\sqrt{eH}X_2\right)e^{-\frac{1}{2}eHX_2^2},\label{eqII8}
\end{eqnarray}
with $H_l$ being the Hermite polynomials and $X_2=\left(x_2-\frac{p_1}{eH}\right)$. From Eq.(\ref{eqII7}) the energy eingenvalues can be obtained, yielding the dispersion relation,
\begin{eqnarray}
(p_0-g_{\omega}\langle\omega_0\rangle)^2=p_3^2+M_{*}^2+\frac{eH}{2}\left(2l+1-s\right),\label{eqII9}
\end{eqnarray}
where $ eH$ is the so-called cyclotron frequency; $s=\pm1$ is the spin variable; and $l=0,1,2,...$, denotes the so-called Landau levels.

Besides, we mention another fundamental consequence of presence of magnetic background. The Feynman rules are modified by the dimensional reduction of four-momentum integral:
\begin{eqnarray}
\int\frac{d^4p}{(2\pi)^4}f(p)\rightarrow\frac{eH}{2\pi}\sum_{l=0}^{\infty}\int\frac{d^2p}{(2\pi)^2}f(p_0,p_3,l).\label{eqII10}
\end{eqnarray}

In order to analyze the thermodynamic behaviour of the system at thermodynamic equilibrium, let us introduce the partition function via the framework of the imaginary time (Matsubara) formalism~\cite{3mats1}, 
\begin{eqnarray}
Z\propto\int\mathcal{D}\psi^{\dagger}\mathcal{D}\psi \; exp\left\{\left[\int_0^{\beta}d\tau 
\int d^3x(\mathcal{L}_E+\mu j_0)
\right]\right\},\label{eqII11}
\end{eqnarray}
where $\beta=1/T$ is the inverse of temperature, $\mu $ is the chemical potential and $\mathcal{L}_E$ is the Lagrangian density given by Eq.~(\ref{eqII1}) in Euclidean space and in mean-field approximation.

To take into account finite-temperature and finite-size effects on the phase structure of the model, we denote the Euclidean coordinate vectors by $x_E=(x_0,x_1,x_2,x_3)$, where $x_0\in[0,\beta]$ and $x_3\in[0,L]$, with $L$ being the size of the compactified spatial dimension, which can be interpreted as the thickness of the system.  As a consequence, the arguments of the function $f(p_0,p_3,l)$ in Eq.~(\ref{eqII10}) must be replaced in the way of generalized Matsubara prescription, i.e.
\begin{eqnarray}
\int\frac{d^2p}{(2\pi)^2}f(p_0,p_3,l)\rightarrow\frac{1}{\beta L}\sum_{n,m=-\infty}^{\infty}f(\omega_n,\omega_m,l),\label{eqII12}
\end{eqnarray}
such that
\begin{eqnarray}
p_0&\rightarrow&\omega_n=\frac{2\pi}{\beta}\left(n+ \frac{1}{2} \right),\nonumber \\
p_3&\rightarrow&\omega_m=\frac{2\pi}{L}\left(m+c\right).
\label{percond}
\end{eqnarray}
Due to the KMS conditions~\cite{Bellac,Kapusta,PR2014}, the fermionic nature of the system imposes an antiperiodic condition in imaginary time coordinate. Notice, however, that there are no restrictions with respect to the periodicity of the spatial compactified coordinate $x_3$. So, the parameter $c$ in Eq.~(\ref{percond}) can assume the values 0 or $1/2$, depending on the physical interest. 

Now we are able to perform the integration over the fields $\psi$ and $\psi^{\dagger}$ in the definition of partition function shown in Eq.~(\ref{eqII11}). Then, keeping in mind the thermodynamic relations and  the modified integration over momentum as in Eq. (\ref{eqII12}), we obtain the thermodynamic potential, 
%\begin{widetext}
\begin{eqnarray}
\frac{U(T,\mu,H,L)}{V}&=& \tilde{U}  - \frac{eH}{\pi\beta L }\sum_{s=\pm1}\sum_{n,m= -\infty}^{\infty}\sum_{l=0}^{\infty}  \nonumber \\
& & \times \ln\left[\left(\omega_{n} - i \mu_{*} \right)^2 + \omega^2_{m}+M_l^2\right], \label{eqII14}
\end{eqnarray}
%\end{widetext}
where
\begin{eqnarray}
 \tilde{U}  & = & \frac{1}{2}m_{\sigma}^2\langle\sigma\rangle^2-\frac{1}{2}m_{\omega}^2\langle\omega_0 \rangle^2, \nonumber \\
M_l& = & \sqrt{M^2_{*}+eH(2l+1-s)},
\label{eqII15} 
\end{eqnarray}
and $\mu_{*}$ is the effective chemical potential related to the fermion field $\psi$, defined by 
\begin{eqnarray}
\mu_{*}=\mu-g_{\omega}\langle\omega ^0 \rangle.\label{eq117}
\end{eqnarray}

In the present work the thermodynamic potential and the gap equations will be treated in the zeta-function regularization approach~\cite{EE}. In this sense, the thermodynamic potential in Eq.~(\ref{eqII15}) can be rewritten as 
\begin{eqnarray}
\frac{U(T,\mu,H,L)}{V}=  \tilde{U}  - \frac{eH}{\beta L}\sum_{s=\pm1}Y^{'}(0),\label{eqII16}
\end{eqnarray}
where $Y(a)$ is the Epstein-Hurwitz inhomogeneous zeta-function, defined by
\begin{eqnarray}
Y(a)&=&\sum_{l=0}^{\infty}\sum_{n,m=-\infty}^{\infty}\left[\omega^2_{n}+\omega^2_{m}+M_l^2\right]^{-a},\label{eqII17}
\end{eqnarray}
with $Y^{'}(a)$ denoting the derivative of $Y(a)$ with respect to the argument $a$. 

One of the advantages of this method is that Epstein-Hurwitz inhomogeneous zeta-function in Eq. (\ref{eqII16}) provides a relatively simple and more tractable way to perform the analytical continuation to the whole complex $a$ plane~\cite{EE,Abreu:2006,Abreu:2009zz}. It has therefore the following representation,
\begin{eqnarray}
Y(a)&=&\frac{\beta L}{\pi}\Bigg[\frac{1}{8}\frac{\Gamma(a-1)}{\Gamma(a)}F_1(a-1)+\frac{1}{\Gamma(a)}F_2(a-1)+\nonumber\\
    & &+\frac{1}{\Gamma(a)}F_3(a-1)+\frac{2}{\Gamma(a)}F_4(a-1)\Bigg],\label{eqII18}
\end{eqnarray}
where the functions $F_1(\nu)$, $F_2(\nu)$, $F_3(\nu)$ and $F_4(\nu)$ are, respectively,
\begin{eqnarray}
F_1(\nu)=(2eH)^{-\nu}\zeta\left(\nu,\frac{1}{2}+\frac{M_{*}^2}{2eH}\right),\label{eqII19}
\end{eqnarray}
\begin{eqnarray}
F_2(\nu)=\sum_{l=0}^{\infty}\sum_{n=1}^{\infty}a_n^{\nu}cosh(n\mu^{*}\beta)\left(\frac{n\beta}{M_l}\right)^{\nu}K_{\nu}(n\beta M_l),\label{eqII20a}
\end{eqnarray}
\begin{eqnarray}
F_3(\nu)=\sum_{l=0}^{\infty}\sum_{m=1}^{\infty}a_m^{\nu}\left(\frac{mL}{M_l}\right)^{\nu}K_{\nu}(m\beta M_l),\label{eqII21}
\end{eqnarray}
and
\begin{eqnarray}
F_4(\nu)=\sum_{l=0}^{\infty}\sum_{n,m=1}^{\infty}b_{n,m}^{\nu}cosh(n\mu^{*}\beta)\times\nonumber\\
        \times\left(\frac{\sqrt{n^2\beta^2+m^2L^2}}{M_l}\right)^{\nu}K_{\nu}(M_l\sqrt{n^2\beta^2+m^2L^2}),\label{eqII22}
\end{eqnarray}
with $\zeta(\eta,a)=\sum_{k=0}^{\infty}(k+a)^{-\eta}$ being the Hurwitz zeta function, $K_{\nu}(z)$ the modified Bessel function, $a_i^{\nu}=\frac{(-1)^i}{2^{\nu}}$ $(i=n,m)$, and $b_{n,m}^{\nu}=2^{\nu}a_n\cdot a_m$.

Looking at the thermodynamic potential in Eq. (\ref{eqII16}) in more detail, the derivative of $Y(a)$ with respect to $a$ at $a\rightarrow 0$ must be carried out cautiously, with a careful analysis of the pole structure~\cite{EE,Abreu:2006,Abreu:2009zz}. Nevertheless,  the following properties of the functions $F_i (i=1,...,4)$ given in Eq.~(\ref{eqII18}), 
\ben
\frac{d}{d\eta}\left[\frac{\Gamma(\eta-1)}{\Gamma(\eta)}F_1(\eta-1)\right]_{\eta\rightarrow\epsilon} & \approx & -(1+\epsilon)F_1^{'}(\epsilon-1) \nonumber \\ 
& & - F_1(\epsilon-1), \nonumber \\
\frac{d}{d\eta}\left[\frac{1}{\Gamma(\eta)}F_b(\eta-1)\right]_{\eta\rightarrow\epsilon} & \approx & F_b(\epsilon-1);\hspace{0.3cm} b=2,3,4,
\nonumber \\ 
\label{pole1}
\een
allow to rewrite the thermodynamic potential as, 
\begin{eqnarray}
\frac{U(T,\mu,L,H)}{V}&=&  \tilde{U} + U_{vac}+\frac{(eH)^2}{4\pi^2}F_5\left(\frac{M_{*}^2}{2eH}\right)-\nonumber\\
                      & &\frac{2eH}{\pi^2}\sum_{s=\pm1}\sum_{i=2}^4(1+\delta_{i,4})F_i(-1),\label{eqII20}
\end{eqnarray}
where 
\begin{eqnarray}
F_5(x)=\frac{\partial\zeta(\eta,x)}{\partial\eta}\bigg|_{\eta=-1}-\frac{1}{2}(x^2-x)\ln(x)+\frac{1}{4}x^2,\label{eq320}
\end{eqnarray}
and $U_{vac}$ is associated to the vacuum fluctuation energy density, i.e. the quantum correction coming from the first term of Eq.~(\ref{eqII18}). As discussed in Refs.~\cite{Chin,Freedman1,MATSUI,PhysRevC.96.055204}, it can be interpreted in the following way: the scalar interaction effectively changes the fermion mass according to Eq.~(\ref{massa}), from $m_{\psi}$ to $M_{*} = m_{\psi} - g_{\sigma} \langle\sigma\rangle$, inducing an energy density shift of the vacuum. However, we remark that our interest is in the finite-size effects on the phase structure of the Walecka's mean-field theory introduced in Ref.~\cite{JTheis:1983PRD} in the presence of a magnetic background, without quantum correction. 
Therefore, hereafter we will omit the term $U_{vac}$ in the calculations, and postpone for a further work the analysis of quantum correction contributions.

The investigation of thermodynamic behaviour also requires the study of the gap  equations, defined as 
\begin{eqnarray}
\frac{\partial U}{\partial\langle\sigma\rangle}&=&0,\label{eq1161}\\
\frac{\partial U}{\partial\langle\omega^{0}\rangle}&=&0.\label{eq1162}
\end{eqnarray}
The solutions of these equations yield the values for $\langle\sigma\rangle$ and $\left\langle\omega^0\right\rangle$  corresponding to extrema of thermodynamic potential, and their use in Eqs.~(\ref{massa}) and~(\ref{eq117}) afford the allowed values of $(T,\mu,L,H)-$dependent effective fermion mass $M_{*}$  and  effective chemical potential $\mu_{*}$. Thus, the substitution of   
Eq.~(\ref{eqII20}) in (\ref{eq1161}) and (\ref{eq1162}) engenders the gap equations rewritten as
\begin{eqnarray}
\left\langle\sigma\right\rangle & = & \frac{g_{\sigma}}{m_{\sigma}^2}\rho_{s}^{H},\label{eq131} \\
\left\langle\omega^0\right\rangle & = & -\frac{g_{\omega}}{m_{\omega}^2} \rho^{H},\label{eq132}
\end{eqnarray}
where the scalar and number densities are written as 
%\begin{widetext}
\begin{eqnarray}
\rho_s^{H} & = & -\frac{eH}{2\pi^2}\left[\frac{1}{4}F_6\left(\frac{M_{*}^2}{2eH}\right)-\sum_{s=\pm1} \sum_{i=2}^4(1+\delta_{i,4})F_i(0)\right], \nonumber \\\label{eq133} \\
\rho^{H} & = & -\frac{eH}{\pi^2} \sum_{s=\pm1}\sum_{l=0}^{\infty}\sum_{n=1}^{\infty}a_nM_l\sinh(n\beta\mu_{*})\bigg[K_1(n\beta M_l)+\nonumber\\
   & &      +\sum_{m=1}^{\infty}a_m\left(\frac{n\beta}{\sqrt{n^2\beta^2+m^2L^2}}\right) \nonumber \\
   & & \times K_1(M_l\sqrt{n^2\beta^2+m^2L^2})\bigg],\label{eq134}
\end{eqnarray}
respectively, with the definition
%\end{widetext} 
\begin{eqnarray}
F_6(x)=\ln\frac{\Gamma(x)}{\sqrt{2\pi}}-\frac{1}{2}(2x-1)\ln(x)+x. \label{eq135}
\end{eqnarray}

From Eq.~(\ref{eq134}), it can be noticed that when the system is considered in effective chemical equilibrium, i.e. $\mu_{*}=0$, the number density acquires a vanishing value, $\rho^{H}=0$, and therefore Eq.~(\ref{eq132}) engenders $\left\langle\omega^0\right\rangle=0$.  

Hence, we have obtained above $(T,\mu,L,H)-$dependent expressions for the thermodynamic potential and scalar and number densities. Another thermodynamic quantities as pressure, entropy and others can be obtained by similar procedures to those described above.

In next section, we will discuss the thermodynamic behaviour of the present system, in presence of an external uniform magnetic field and boundaries.

%%%%%%%%%%%%%%%%%%%%%%%%%%%%%%%%%%%%%%%%%%%%%%%%%%%%%%%%%%%%%%%%%%%%%%%%%%%%%%%%%%%%%%%
%%%%%%%%%%%%%%%%%%%%%%%%%%%%%%%%%%%%%%%%%%%%%%%%%%%%%%%%%%%%%%%%%%%%%%%%%%%%%%%%%%%%%%%%
\section{Phase Structure and Comments}
%%%%%%%%%%%%%%%%%%%%%%%%%%%%%%%%%%%%%%%%%%%%%%%%%%%%%%%%%%%%%%%%%%%%%%%%%%%%%%%%%%%%%%%
%%%%%%%%%%%%%%%%%%%%%%%%%%%%%%%%%%%%%%%%%%%%%%%%%%%%%%%%%%%%%%%%%%%%%%%%%%%%%%%%%%%%%%%%

This section is devoted to the analysis of the phase structure of the system, focusing on how it behaves with the change of the relevant parameters $(T,\mu,L,H)$ of the model, and in special, the influence of the magnetic background and boundaries. In the present investigation the system can be regarded as a simplified model to describe the nuclear matter in a medium, also considered at effective chemical equilibrium, i.e. $\mu_{*}=0$. As discussed in previous section, this fact engenders a vanishing solution $\left\langle\omega^0\right\rangle=0$ of Eq.~(\ref{eq132}). 
In other words, it means that the physical system has the same number of fermion and antifermions. 
%Thus, we focus on the thermodynamics of the system as a function of the gap equation for $\langle\sigma\rangle=\langle\sigma\rangle(T,L,\Omega)$ or $M_{*}=M_{*}(T,L,\Omega)$ at $\left\langle\omega^0\right\rangle=0$.
%

%It is appropriate to scale all physical quantities in terms of units of mass of the fermion field  $m_{\psi}$, i.e.   
%
%\begin{eqnarray}
%
%\frac{U}{m_{\psi}^4} & \rightarrow &  U, \;\; \frac{T}{m_{\psi}}\rightarrow T,\;\; \frac{\mu}{m_{\psi}}
%
%\rightarrow\mu, \;\; \frac{\sigma}{m_{\psi}}\rightarrow\sigma, \nonumber \\
%
%\frac{m_{\sigma}}{m_{\psi}} &  \rightarrow & m_{\sigma},\;\; \frac{M_{*}}{m_{\psi}}\rightarrow M_{*},\;\; Lm_{\psi}\rightarrow L , \;\; \frac{\Omega}{m_{\psi}^2}\rightarrow\Omega, 
%
%\label{eqIII1}
%
%\end{eqnarray}
%
%where $\Omega = e H $ denotes the so-called cyclotron frequency. Therefore, henceforth the parameter $\Omega $ will be associated to the magnitude of magnetic field.  

The numerical parameters of the Walecka model we use are according to the scenario of hadronic physics, in which the fermionic field is associated to nucleons and scalar and vector fields are associated to isoscalar-scalar
$(\sigma)$ and isoscalar-vector $(\omega)$ mesons. In this sense, The values of parameters are~\cite{Ishikawa:2018yey,Buballa:1996tm}: $m_{\psi} = 939$ MeV, $m_{\sigma} = 500$ MeV, $m_{\omega} = 783$ MeV. Concerning the value of coupling constant $g_{\sigma}$, we discuss in next subsection.

%The scenario presented in the equations of the previous section reduces the system to a confinement between two parallel planes at a distance $L$ from each other.
%

%%%%%%%%%%%%%%%%%%%%%%%%%%%%%%%%%%%%%%%%%%%%%%%%%%%%%%%%%%%%%%%%%%%%%%%%%%%%%%%%%%%%%%%
%%%%%%%%%%%%%%%%%%%%%%%%%%%%%%%%%%%%%%%%%%%%%%%%%%%%%%%%%%%%%%%%%%%%%%%%%%%%%%%%%%%%%%%%
\subsection{System without spatial boundaries}
%%%%%%%%%%%%%%%%%%%%%%%%%%%%%%%%%%%%%%%%%%%%%%%%%%%%%%%%%%%%%%%%%%%%%%%%%%%%%%%%%%%%%%%
%%%%%%%%%%%%%%%%%%%%%%%%%%%%%%%%%%%%%%%%%%%%%%%%%%%%%%%%%%%%%%%%%%%%%%%%%%%%%%%%%%%%%%%%

We start, for completeness, by studying the behaviour of the nucleon effective mass $M_{*}$ under changes of parameters but without the presence of boundaries, which is basically the scenario described in Ref.~\cite{JTheis:1983PRD} but with the presence of magnetic background~\footnote{Notice that we use a different scaling and notation with respect to Ref.~\cite{JTheis:1983PRD}.}. 
The gap equation for $\sigma$-field in Eq.~(\ref{eq131}) gives the expected value of the field $\sigma$, and its use in Eq.~(\ref{massa}) yields the nucleon effective mass $M_{*}$ of the field $\psi$ in medium as function of relevant parameters. 
%In what follows all parameters are understood to be redefined by the scaling in Eq.~(\ref{eqIII1}). 
%

It is worthy mentioning that in the situation without the presence of external magnetic field, Refs.~\cite{JTheis:1983PRD,PhysRevC.96.055204} have reported the influence of the magnitude of the coupling constant  $g_{\sigma} $ on the nature of the phase diagram: for $g_{\sigma} < 9.8 $ the nucleon effective mass is  smooth in the temperature, whereas for $g_{\sigma} > 9.8 $ a phase transition of first order takes place. Therefore, we estimate the effect of a magnetic background on this property by analyzing in the next two figures the phase structure of the system in the situations of lower and greater values of $g_{\sigma} $.

\begin{figure}[th]
\centering
\includegraphics[{height=8.0cm,width=8.0cm}]{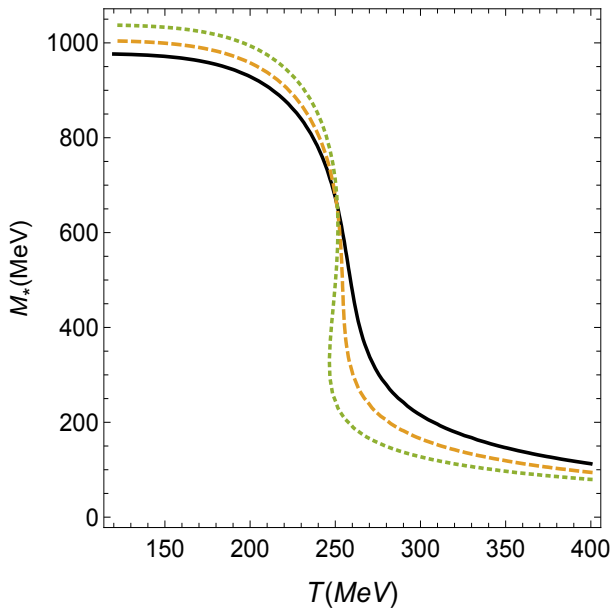}
\caption{ Plot of nucleon effective mass in Eq.(\ref{eq131}) as a function of temperature, at chemical equilibrium. We fix $g_{\sigma}=5.0$, with full, dashed and dotted lines representing respectively $\Omega = 3 \times 10^{5}, 4 \times 10^{5}$, and $5 \times 10^{5}$ MeV${}^{2}$ at $L\rightarrow\infty$.} 
\label{fig:MassaEfetivaMC0}
\end{figure}

In Fig.~\ref{fig:MassaEfetivaMC0} is plotted the values of $M_{*}$ that are solutions of the gap equation in Eq. (\ref{eq131}) as function of temperature for different values of  $\Omega $,  keeping the values of $g_{\sigma} < 9.8 $ and $m_{\sigma} $ fixed. We can observe that as the temperature increases, the nucleon effective mass acquires small values, which means that the 
system decouples to like an almost-free zero-mass fermion gas. But the point here is the influence of magnetic background: at zero temperature, the augmentation of $\Omega$ enhances the broken phase (i.e. the magnetic catalysis effect~\cite{Fraga}). This phenomenon remains up to a certain value of temperature.  However, an opposite behaviour appears at higher temperatures, where the system tends toward symmetric phase faster as the field strength increases. In other words, an inverse magnetic catalysis takes places in this regime.  This phenomenon is known in the literature, and also found in other scenarios~\cite{MAO,Tobias,Mamo:2015dea,Pagura,Magdy,Ayala2,Ayala0}. This mechanism induces the restoration of  symmetry at higher temperatures of the system, lost at lower temperatures. 

Besides, another relevant feature from the behaviour of $M_{*}$  with $\Omega$ in Fig.~\ref{fig:MassaEfetivaMC0} is that the nature of the transition is modified with the growth of magnetic field: a smooth transition gives rise to a discontinuous one.  
Namely, there is a value for $\Omega$ above which the system suffers a sudden transition (about $5 \times 10^{5}$ MeV${}^{2}$ for $g_{\sigma}=5.0$). 
%
%where the dependence of $M_{*}$ on $T$ is continuous \cite{}, for different values 
%
%magnetic field and with larger $g_{\sigma}$, for 
%
%different values of the magnetic field.
%

\begin{figure}[th]
\centering
\includegraphics[{height=8.0cm,width=8.0cm}]{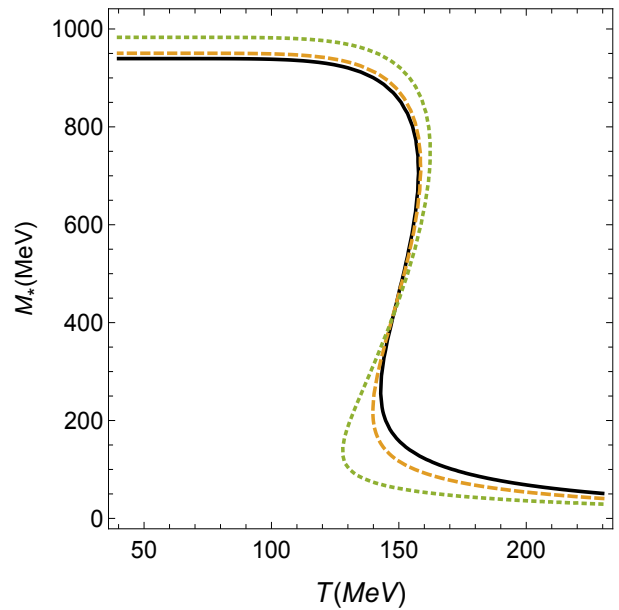}
\caption{  Plot of nucleon effective mass in Eq.(\ref{eq131}) as a function of temperature, at chemical equilibrium. We fix $g_{\sigma}=16.00$, with full, dashed and dotted lines representing respectively $\Omega \approx 1 \times 10^{4}, 5 \times 10^{4}$, and $1 \times 10^{5}$ MeV${}^{2}$ at $L\rightarrow\infty$.} 
\label{fig:MassaEfetivaMC1}
\end{figure}

In Fig.~\ref{fig:MassaEfetivaMC1} is shown the plot similar to the one in Fig. \ref{fig:MassaEfetivaMC0}, but keeping $g_{\sigma} > 9.8 $ and $m_{\sigma} $ fixed. It can be remarked that this regime of greater magnitude of attractive interaction gives rise to S-shaped curves with mixed phase regions, characterizing a discontinuous phase transition at smaller critical temperatures than the previous case.
Also, the increase of magnetic field favours the symmetric phase, with effective mass suddenly decreasing at lower critical temperatures. This behaviour is similar to a liquid-gas transition, but is opposite to the one found in the bosonic context~\cite{PhysRevC.96.055204}. As in Fig.~\ref{fig:MassaEfetivaMC0}, magnetic catalysis (inverse magnetic catalysis) happens at smaller (higher) temperatures.

To better characterize the first-order nature of the transition, the global minimum and transition temperature, one must analyze the 
thermodynamic potential density at temperatures where the curves  of $M_{\ast} \times T$ are S-shaped.  We use the following normalization for $U$ with respect to reference $U(M_{*}=0)$:
\begin{eqnarray}
\frac{ \overline{U}}{V} \equiv \frac{1}{V}[U(M_{*})-U(M_{*}=0)].
\label{norm_pot}
\end{eqnarray}

\begin{figure}[th]
\centering
\includegraphics[{height=8.0cm,width=8.0cm}]{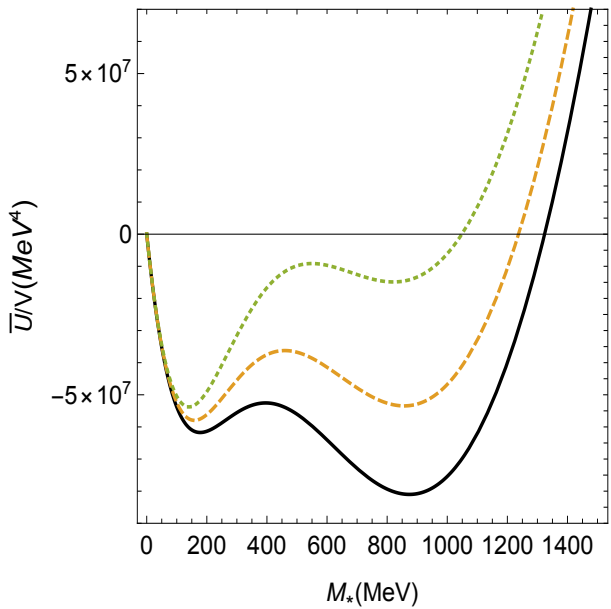}
\caption{  Plot of thermodynamic potential density in Eq.~(\ref{norm_pot}) as a function of nucleon effective mass, at chemical equilibrium, for $g_{\sigma}=16.0$, $\Omega = 1 \times 10^{4}$ MeV${}^{2}$ and $L\rightarrow\infty$. Full, dashed and dotted lines represent the cases for $T = 147, 150$ and $T=154$ MeV, respectively.} 
\label{fig:PotencialMC1}
\end{figure}

\begin{figure}[th]
\centering
\includegraphics[{height=8.0cm,width=8.0cm}]{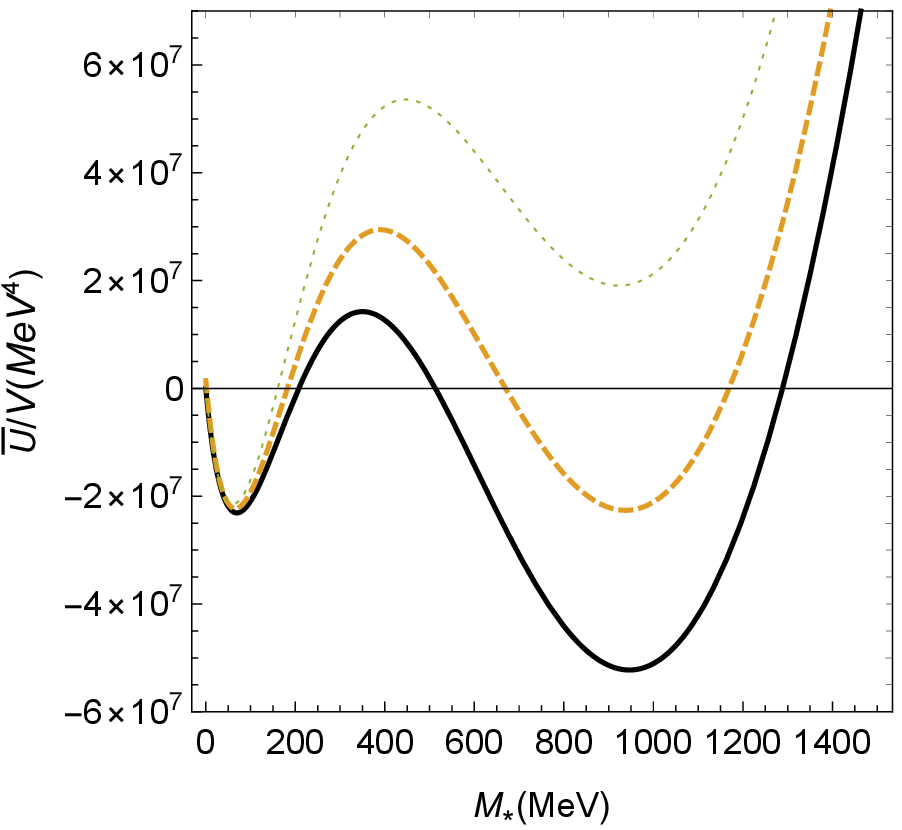}
\caption{ Plot of thermodynamic potential density in Eq.~(\ref{norm_pot}) as a function of nucleon effective mass, at chemical equilibrium, for $g_{\sigma}=16.0$, $\Omega \approx 1 \times 10^{5}$ MeV${}^{2}$  and $L\rightarrow\infty$. Full, dashed and dotted lines represent the cases for $T=143, 146$ and $150$ MeV, respectively.} 
\label{fig:PotencialMC2}
\end{figure}

\begin{figure}
\centering
\includegraphics[{height=8.0cm,width=8.0cm}]{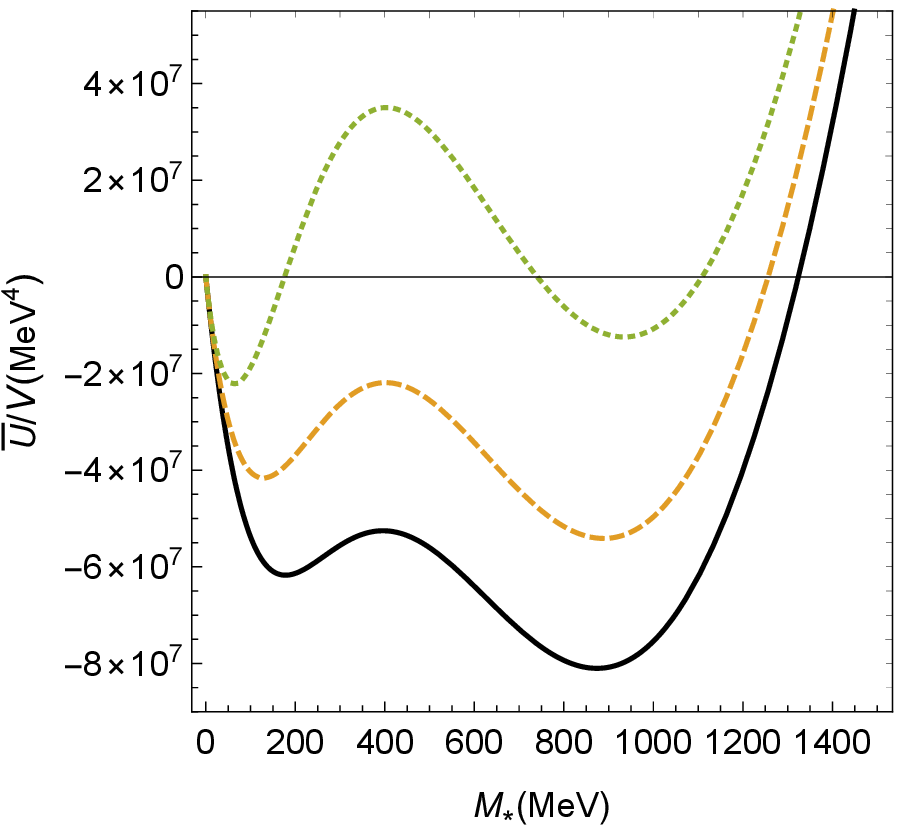}
\caption{Plot of thermodynamic potential density in Eq.~(\ref{norm_pot}) as a function of nucleon effective mass, at chemical equilibrium, for $g_{\sigma}=16.0$ and $T =v147$ MeV and $L\rightarrow\infty$. Full, dashed and dotted lines represent the cases for $\Omega = 1 \times 10^{4}, 5 \times 10^{4}$, and $1 \times 10^{5}$ MeV${}^{2}$, respectively.} 
\label{fig:PotencialMC3}
\end{figure}

Thus, in Figs.~\ref{fig:PotencialMC1}-\ref{fig:PotencialMC2} the normalized thermodynamic potential density $\overline{U} / V $, obtained by using Eq.~(\ref{eqII20}) in (\ref{norm_pot}), is plotted  as a function of nucleon effective mass taking respectively two different values of $\Omega $, with distinct temperatures in the transition range, but with $m_{\sigma}$, $g_{\sigma} > 9.8$ kept fixed. As suggested in Fig.~\ref{fig:MassaEfetivaMC1}, a two-step phase transition occurs as the temperature increases. At lower temperatures, the global minimum is at a higher values of $M_{*}$; the increase of $T$ makes the second local minimum in small values of $M_{*}$ overcome the former one, 
becoming global minimum; and for higher temperatures the absolute minimum tends smoothly toward zero. Besides,  as the field strength increases, the first-order phase transition occurs at smaller critical temperatures, with the global minimum approaching to zero faster.

In Fig. \ref{fig:PotencialMC3} we plot $\overline{U} / V $ as a function of effective mass, at different values of magnetic field, but with temperature,  $m_{\sigma}$ and $g_{\sigma}> 9.8$ kept fixed. It can be noticed in more detail the effect of inverse magnetic catalysis discussed above: in this range of temperature the augmentation of $\Omega $ induces the restoration of symmetry at higher temperatures of the system, lost at lower temperatures.

%%%%%%%%%%%%%%%%%%%%%%%%%%%%%%%%%%%%%%%%%%%%%%%%%%%%%%%%%%%%%%%%%%%%%%%%%%%%%%%%%%%%%%%%
%  Referee - point 2 - LL 
%%%%%%%%%%%%%%%%%%%%%%%%%%%%%%%%%%%%%%%%%%%%%%%%%%%%%%%%%%%%%%%%%%%%%%%%%%%%%%%%%%%%%%%%
We also analyze the properties of the system concerning the magnetic field strength by considering the filling of Landau levels (LL). 
The number of occupied Landau levels $(l_{max})$ is examined by neglecting all levels that contribute with an amount which yields an error less than $0.1\%$ in effective mass. We see from Table~\ref{table1} that the number of LL increases when the temperature grows at fixed magnetic field strength. On the other hand, at fixed temperature the number of LL decreases as the magnetic field raises. It is in accordance with the results available in existing literature~\cite{PhysRevC.83.065805}. This is relevant due to the fact that a smaller number of occupied LL makes easier the numerical computation of the results. 

\begin{center}
\begin{table}[h!]
\caption{   Number of occupied Landau levels $(l_{max})$  that contributes with an amount which yields an error less than $0.1\%$ in effective mass.}
\vskip1.5mm
\label{table1}
\begin{tabular}{c | c  | c }
\hline
\hline
$\Omega$ (MeV${}^2$) & $T$ (MeV)  &  $l_{max}$ \\
\hline
$1 \times 10^4$ &  50 &  1
\\
$1 \times 10^5$ &  50  & 1
\\
$1 \times 10^4$ & 140  & 15
\\
$1 \times 10^5$ & 140  & 4
\\
$1 \times 10^4$ & 160  & 42
\\
$1 \times 10^5$ & 160  & 20
%\\
%$1 \times 10^4$ & 200  & 55
%\\
%$1 \times 10^5$ & 200  & 5
\\
\hline
\hline
\end{tabular}
\end{table}
\end{center}

%%%%%%%%%%%%%%%%%%%%%%%%%%%%%%%%%%%%%%%%%%%%%%%%%%%%%%%%%%%%%%%%%%%%%%%%%%%%%%%%%%%%%%%%
%%%%%%%%%%%%%%%%%%%%%%%%%%%%%%%%%%%%%%%%%%%%%%%%%%%%%%%%%%%%%%%%%%%%%%%%%%%%%%%%%%%%%%%% Point 3, 5 of referee

\begin{figure}
\centering
\includegraphics[{height=8.0cm,width=8.0cm}]{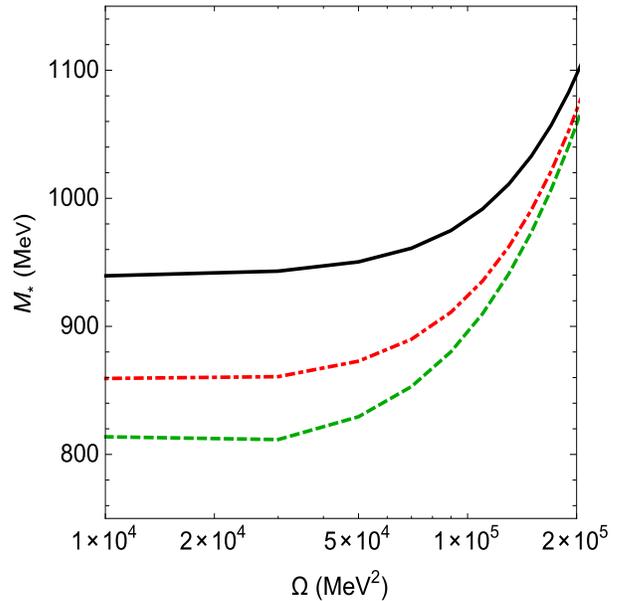}
\caption{  Plot of nucleon effective mass in Eq.(\ref{eq131}) as a function of $\Omega$, at chemical equilibrium, for $g_{\sigma}=16.0$ and $L\rightarrow\infty$. Full, dotdashed and dashed lines representing respectively $T = 50, 100$, and $155$ MeV at $L\rightarrow\infty$.} 
\label{fig:PotencialMC4}
\end{figure}

We complete this subsection by examining in more detail the influence of the magnetic field and temperature on the effective mass, as well as their typical ranges where the system is more susceptible to a phase transition. In Fig.~\ref{fig:PotencialMC4} we display $M_{\ast}$ in terms of the magnetic field strength $\Omega$, for three different values of temperature. It can be seen that the system becomes sensitive to the effects of magnetic field for magnitudes higher than $3 \times 10^4$ MeV${}^2$. The competing character of temperature and magnetic field effects is evident in the considered ranges: while the increase of temperature induces the restoration of chiral symmetry, the augmentation of magnetic field strength yields an opposite effect.  Besides, when qualitatively compared to the results in the context of Nambu-Jona-Lasinio model~\cite{PhysRevC.83.065805,Grunfeld:2014qfa}, our outcomes present noticeable similarities.

%%%%%%%%%%%%%%%%%%%%%%%%%%%%%%%%%%%%%%%%%%%%%%%%%%%%%%%%%%%%%%%%%%%%%%%%%%%%%%%%%%%%%%%%
%%%%%%%%%%%%%%%%%%%%%%%%%%%%%%%%%%%%%%%%%%%%%%%%%%%%%%%%%%%%%%%%%%%%%%%%%%%%%%%%%%%%%%%%
\subsection{System with compactified spatial dimensions}
%%%%%%%%%%%%%%%%%%%%%%%%%%%%%%%%%%%%%%%%%%%%%%%%%%%%%%%%%%%%%%%%%%%%%%%%%%%%%%%%%%%%%%%%
%%%%%%%%%%%%%%%%%%%%%%%%%%%%%%%%%%%%%%%%%%%%%%%%%%%%%%%%%%%%%%%%%%%%%%%%%%%%%%%%%%%%%%%%

Here we investigate the influence of boundaries on the phase structure. It means that the system exists in a region delimited by two infinite planes at a finite distance $L$ from each other. We concentrate on the situation of larger values of $ g_{\sigma} $, in which a phase transition of first order takes place. Some of the features of finite-size effects have also discussed in Ref.~\cite{PhysRevC.96.055204}.

\begin{figure}[th]
\centering
\includegraphics[{height=8.0cm,width=8.0cm}]{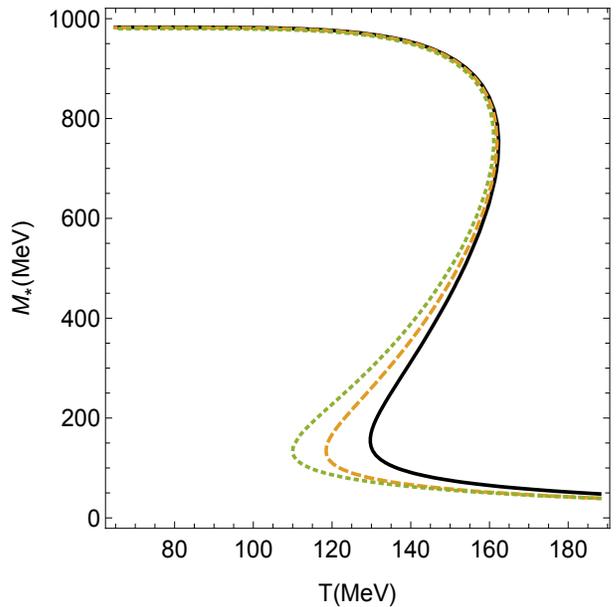}
\caption{Plot of nucleon effective mass in Eq.(\ref{eq131}) as a function of temperature, at chemical equilibrium. We fix $g_{\sigma}=16.0$. Full, dashed and dotted lines represent $L = 5, 2$ and $1.8$ fm, respectively, at  $\Omega = 1 \times 10^{5}$ MeV${}^{2}$.} 
\label{fig:MassaEfetivaMCX1}
\end{figure}

The plot in Fig.~\ref{fig:MassaEfetivaMCX1} is the same as in Fig.~\ref{fig:MassaEfetivaMC0}, but for finite values of $L$ and with $\Omega$ kept fixed. We see that allowed values of effective mass  are affected by the presence of boundaries; the range of temperature where occurs the mixed phase is spread out as the length of compactified coordinate decreases. In other words, the symmetric phase is favoured as the size of the system decreases. 

\begin{figure}[th]
\centering
\includegraphics[{height=8.0cm,width=8.0cm}]{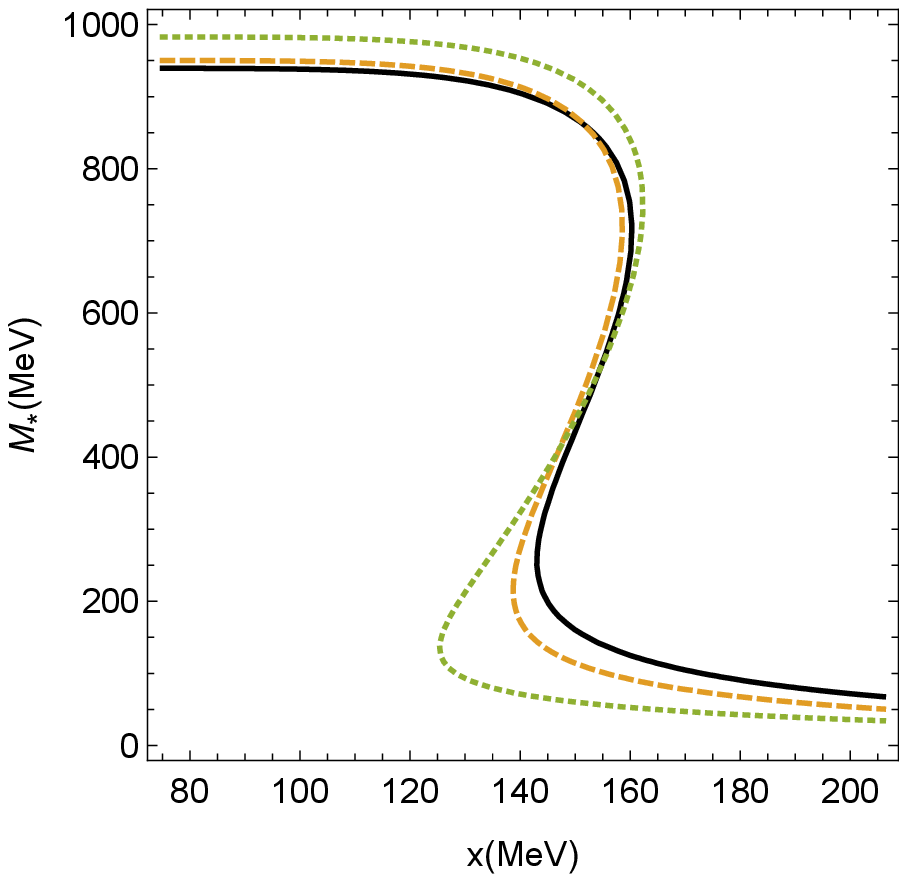}
\caption{ Plot of effective mass in Eq.(\ref{eq131}) as a function of inverse of the length $x=1/L$, at chemical equilibrium. We fix $g_{\sigma}=16.0$ and $T = 80 $ MeV. Full, dashed and dotted lines represent $\Omega \approx 1 \times 10^{4}, 5 \times 10^{4}$, and $1 \times 10^{5}$ MeV${}^{2}$.}
\label{fig:MassaEfetivaMCX2}
\end{figure}

The combined effects of the dependence on the size of the system and field strength can be better described from Fig.~\ref{fig:MassaEfetivaMCX2}, in which 
is plotted the values of effective mass that are solutions of the gap equation in Eq. (\ref{eq131}) as function of inverse of thickness $x=1/L$ for different values of  $\Omega $,  keeping temperature, $g_{\sigma} > 9.8 $ and $m_{\sigma} $ fixed. 
In the bulk ($x\rightarrow 0$ or $L\rightarrow\infty$), $M_{*}$ suffers an increase as field strength grows, as expected due to magnetic catalysis effect previously discussed. Notice that the reduction of $L$ engenders a reduction of $M_{*}$, which is a similar behaviour with respect to the temperature dependence. In this context, the results suggest the existence of a critical value $L_c$ at which the system experiences a discontinuous phase transition to a small value of $M_{*}$. The interesting point here is that the growth of field strength induces greater values for $L_c$, i.e. the increase of $\Omega$ stimulates the abrupt drop of $M_{*}$ at larger values of $L$. 

\begin{figure}
\centering
\includegraphics[{height=8.0cm,width=8.0cm}]{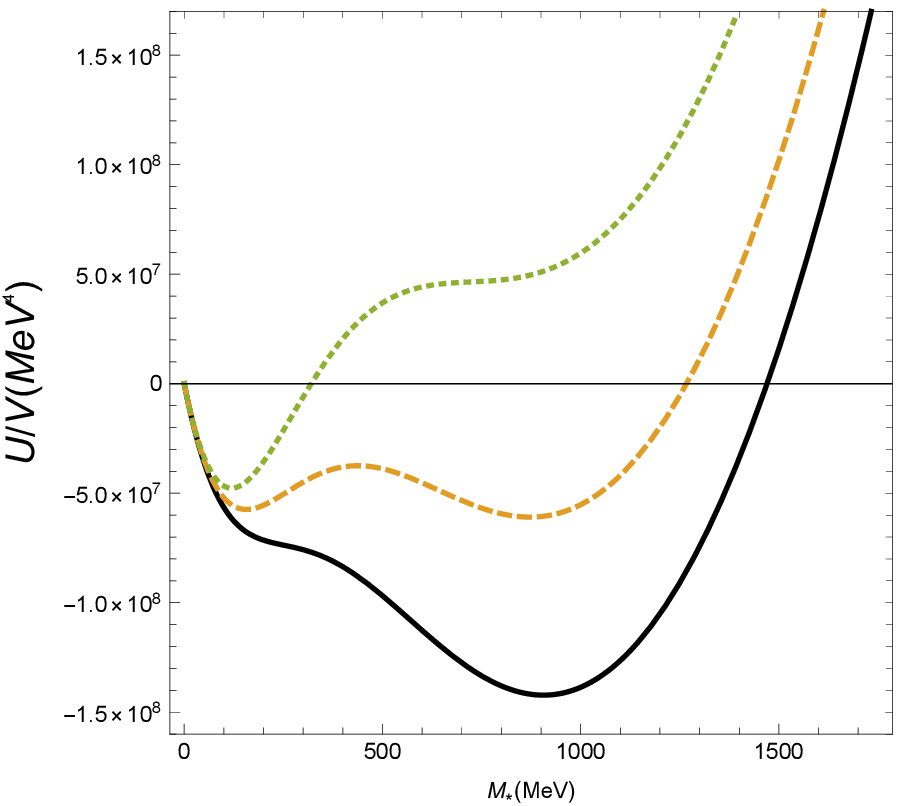}
\caption{ Plot of thermodynamic potential density of Eq.~(\ref{eqII20}) as a function of effective mass, at chemical equilibrium, at  $g_{\sigma}=16.0$, $T=80$ MeV and $\Omega = 1 \times 10^{4} $ MeV${}^{2}$. Full, dashed and dotted lines represent the cases for $L \approx 1.4, 1.3$ and $1.2$ fm, respectively.}
\label{fig:PotencialMCX1}
\end{figure}

\begin{figure}
\centering
\includegraphics[{height=8.0cm,width=8.0cm}]{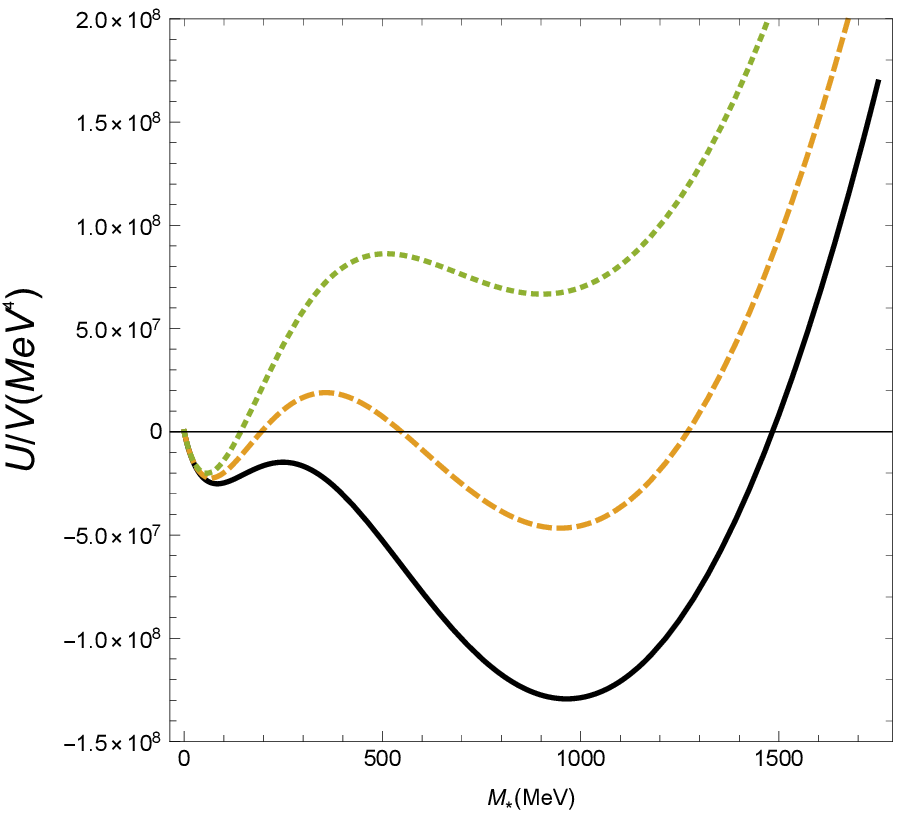}
\caption{Plot of thermodynamic potential density of Eq.~(\ref{eqII20}) as a function of effective mass, at chemical equilibrium, at $g_{\sigma}=16.0$, $T=80$ MeV and $\Omega = 1 \times 10^{5}$ MeV${}^{2}$. Full, dashed and dotted lines represent the cases for $L \approx 1.5, 1.4$ and $1.3$ fm, respectively.}
\label{fig:PotencialMCX2}
\end{figure}

Once more, for the sake of completeness a plot of the thermodynamic potential density $\overline{U} / V $ as a function of effective mass is shown in Figs.~\ref{fig:PotencialMCX1} and \ref{fig:PotencialMCX2}, at different values of $L$ but with field strength, temperature, $m_{\sigma}$ and $g_{\sigma}$ kept fixed. It can be remarked that higher values of $L$ the global minimum is localized at bigger values of $M_{*}$; as the size of the system reduces, at a specific critical value $L_c$ the absolute minimum becomes the one at smaller values of $M_{*}$; even bigger values of $L$  make $M_{*}$ tending smoothly toward zero. Moreover, the first-order phase transition occurs at bigger critical sizes as the field strength grows, with the global minimum moving towards zero faster.

%%%%%%%%%%%%%%%%%%%%%%%%%%%%%%%%%%%%%%%%%%%%%%%%%%%%%%%%%%%%%%%%%%%%%%%%%%%%%%%%%%%%%%%%
%%%%%%%%%%%%%%%%%%%%%%%%%%%%%%%%%%%%%%%%%%%%%%%%%%%%%%%%%%%%%%%%%%%%%%%%%%%%%%%%%%%%%%%%

Hence, our findings suggest that the presence of boundaries disfavours the maintenance of long-range correlations, inducing the suppression of the ordered phase. In the end, we see that the phase structure of the system is strongly affected by the combination of the effects associated to the existence of boundaries and magnetic background.  We can summarize as follows: starting at smaller temperatures, with the system in the broken phase, the bulk approach seems a good approximation in the range of greater values of the thickness $L$, since the magnetic catalysis is not modified. 
The effective mass behavior obtained in the present approach has noticeable similarities with the one in the context of Nambu-Jona-Lasinio model discussed in Refs.~\cite{PhysRevC.83.065805,Grunfeld:2014qfa}. 
Nevertheless, keeping $T$ fixed, the reduction of $L$  engenders a change of behaviour and produces a discontinuous phase transition, whose critical value $L_c$ is larger as the field strength increases. Yet at the same temperature and at a thickness $L < L_c$, the symmetric phase is favoured due to both inverse magnetic catalysis effect and the reduction of $L$. These are the main results of this work.

%%%%%%%%%%%%%%%%%%%%%%%%%%%%%%%%%%%%%%%%%%%%%%%%%%%%%%%%%%%%%%%%%%%%%%%%%%%%%%%%%%%%%%%%
%%%%%%%%%%%%%%%%%%%%%%%%%%%%%%%%%%%%%%%%%%%%%%%%%%%%%%%%%%%%%%%%%%%%%%%%%%%%%%%%%%%%%%%%
\section{CONCLUSIONS}
%%%%%%%%%%%%%%%%%%%%%%%%%%%%%%%%%%%%%%%%%%%%%%%%%%%%%%%%%%%%%%%%%%%%%%%%%%%%%%%%%%%%%%%%
%%%%%%%%%%%%%%%%%%%%%%%%%%%%%%%%%%%%%%%%%%%%%%%%%%%%%%%%%%%%%%%%%%%%%%%%%%%%%%%%%%%%%%%%

In this work we have analyzed the phase structure 
of Walecka model in the presence of a magnetic background and boundaries. In mean-field approximation and at effective chemical equilibrium, we have investigated the thermodynamic potential and gap equation solutions under the change of the size of compactified coordinate, temperature and magnitude of external magnetic field. We have interested on the situation of larger values of the coupling constant $ g_{\sigma} $, in which a phase transition of first order takes place as the temperature increases. 

Looking at the magnetic background influence, distinct phenomena appear for different ranges of temperatures: magnetic catalysis (enhancement of broken phase) occurs at smaller values of $T$, while the inverse magnetic catalysis effect (stimulation of the restoration of symmetry) at higher temperatures is suggested.

When the presence of boundaries is taken into account, the maintenance of long-range correlations is disfavoured, inducing the suppression of the ordered phase.  We have seen that the thermodynamic behaviour is strongly affected by boundaries and magnetic background. Taking smaller temperatures, with the system in the broken phase, we have noticed that magnetic catalysis is not altered in the range of greater values of the thickness $L$ (bulk approximation). On the other hand, keeping $T$ fixed, the reduction of  $L$ engenders a discontinuous phase transition at a critical $L_c$, whose value grows as the field strength increases. Therefore, the  symmetric phase is favoured due to both inverse magnetic catalysis effect and the reduction of $L$.

Finally, we stress that the results outlined above can give us insights about relativistic fermionic systems in a hot medium confined in a reservoir. Further studies will be done in order to apply the present approach to specific physical systems (such as ultrarelativistic nuclear and compact astrophysical objects) and to extend it including quantum corrections as well as beyond mean-field approximations. 

%%%%%%%%%%%%%%%%%%%%%%%%%%%%%%%%%%%%%%%%%%%%%%%%%%%%%%%%%%%%%%%%%%%%%%%%%%%%%%%%%%%%%%%%
%%%%%%%%%%%%%%%%%%%%%%%%%%%%%%%%%%%%%%%%%%%%%%%%%%%%%%%%%%%%%%%%%%%%%%%%%%%%%%%%%%%%%%%%
%\section{ACKNOWLEDGMENTS}
\acknowledgments
%%%%%%%%%%%%%%%%%%%%%%%%%%%%%%%%%%%%%%%%%%%%%%%%%%%%%%%%%%%%%%%%%%%%%%%%%%%%%%%%%%%%%%%%
%%%%%%%%%%%%%%%%%%%%%%%%%%%%%%%%%%%%%%%%%%%%%%%%%%%%%%%%%%%%%%%%%%%%%%%%%%%%%%%%%%%%%%%%
 L.M.A. would like to thank the Brazilian funding agencies CNPq (contracts 308088/2017-4 and 400546/2016-7) and FAPESB (contract INT0007/2016) for partial financial support. E.S.N. acknowledges CAPES (Brazilian agency) for financial support.

\vfill \eject
%\newpage


\begin{thebibliography}{99}

\bibitem{Bellac} M. Le Bellac, \textit{Thermal Field Theory} (Cambridge University 
Press, Cambridge, UK, 1996).

\bibitem{Kapusta} J.I. Kapusta and C. Gale, \textit{Finite-Temperature Field Theory: Principles
and Applications} (Cambridge University Press, Cambridge, UK, 2006). 

\bibitem{Ashok} A. Das, \textit{Finite Temperature Field Theory} (World Scientific, Singapore, 1997).

\bibitem{Walecka:1974qa}
  J.~D.~Walecka,
  %``A Theory of highly condensed matter,''
  Annals Phys.\  {\bf 83}, 491 (1974).
  %%CITATION = APNYA,83,491;%%
  
\bibitem{JTheis:1983PRD}J. Theis, G. Graebner,G. Buchwald, J. Maruhn, W. Greiner,
H. Stocker, J. Polonyi, Phys. Rev. D \textbf{28}, 2286-2290 (1983).

\bibitem{Saito}K. Saito, K. Tsushima, D.H. Lu and A.W. Thomas, Phys. Rev. C \textbf{59}, 1203 (1999).
%29

\bibitem{Menezes0} D. P. Menezes, C. Providencia, M. Chiapparini, M.E. Bracco, A. Delfino and M. Malheiro, Phys. Rev. C \textbf{76}, 064902 (2007).

\bibitem{Delfino}A. Delfino, M. Jansen and V.S. Timoteo, Phys. Rev. C \textbf{78}, 034909 (2008).

\bibitem{Lavagno}A. Lavagno, Phys. Rev. C \textbf{81}, 044909 (2010).
%30
%31

%32
%33
\bibitem{Shao} G.Y. Shao, M. Colonna, M. Di Toro, Y.X. Liu and B. Liu, Phys. Rev. D \textbf{87}, 096012 (2013).
%34
\bibitem{Casali} R. H. Casali, L. B. Castro and D. P. Menezes, Phys. Rev. C \textbf{89}, 015805 (2014).
%35

%36
\bibitem{Fukushima} K. Fukushima and C. Sasaki, Progress in Particle and Nuclear Physics \textbf{72}, 99 (2013).
%37
\bibitem{Dutra} M. Dutra, \textit{et al}., Phys. Rev. C \textbf{90}, 055203 (2014).

\bibitem{Torres2} J.R. Torres, F. Gulminelli and D.P. Menezes, Phys.Rev. C \textbf{93}, 024306 (2016). 


%\cite{Zhang:2017etr}
\bibitem{Zhang:2017etr} 
  Z.~W.~Zhang and L.~W.~Chen,
  %``Low density nuclear matter with light clusters in a generalized nonlinear relativistic mean-field model,''
  Phys.\ Rev.\ C {\bf 95}, 064330 (2017).
%  doi:10.1103/PhysRevC.95.064330
%  [arXiv:1705.00555 [nucl-th]].
  %%CITATION = doi:10.1103/PhysRevC.95.064330;%%

%\cite{Oertel:2016bki}
\bibitem{Oertel:2016bki} 
  M.~Oertel, M.~Hempel, T.~Klahn and S.~Typel,
  %``Equations of state for supernovae and compact stars,''
  Rev.\ Mod.\ Phys.\  {\bf 89}, 015007 (2017)
%  doi:10.1103/RevModPhys.89.015007
%  [arXiv:1610.03361 [astro-ph.HE]].
  %%CITATION = doi:10.1103/RevModPhys.89.015007;%%
  %27 citations counted in INSPIRE as of 10 Jul 2017
 
\bibitem{Kharzeev}D.E Kharzeev, L.D. Mclerran and H.J. Warringa, ArXiv:0711.0950 [hep-ph], 2007.
 
 
\bibitem{Skokov:2009qp}  V.~Skokov, A.~Y.~Illarionov and V.~Toneev, Int.\ J.\ Mod.\ Phys.\ A {\bf 24}, 5925 (2009).  

 %\cite{Chernodub:2010qx}
\bibitem{Chernodub:2010qx} 
  M.~N.~Chernodub,
  %``Superconductivity of QCD vacuum in strong magnetic field,''
  Phys.\ Rev.\ D {\bf 82}, 085011 (2010).
 % doi:10.1103/PhysRevD.82.085011
 % [arXiv:1008.1055 [hep-ph]].
  %%CITATION = doi:10.1103/PhysRevD.82.085011;%%
  %179 citations counted in INSPIRE as of 15 Dec 2018


\bibitem{Ayala1} A. Ayala, M. Loewe, J.C. Rojas and C. Villavicencio, Phys. Rev. D \textbf{86}, 076006 (2012).


 \bibitem{Tobias} M. Ferreira, P. Costa, O. Lourenco, T. Frederico, and C. Providencia,  Phys. Rev. D {\bf 89}, 116011 (2014).
  
\bibitem{Heber} A. Haber, F. Preis and A. Schmitt, Phys. Rev. D \textbf{90}, 125036 (2014).

\bibitem{MAO} S. MAO, Phys. Lett. B {\bf 758}, 195 (2016).

%50
\bibitem{Ayala2} A. Ayala, P. Mercado and C. Villavicencio, Phys. Rev. C \textbf{95}, 014904 (2017).





\bibitem{Mamo:2015dea} 
  K.~A.~Mamo,
  %``Inverse magnetic catalysis in holographic models of QCD,''
  JHEP {\bf 1505}, 121 (2015)
%  doi:10.1007/JHEP05(2015)121
%  [arXiv:1501.03262 [hep-th]].
  %%CITATION = doi:10.1007/JHEP05(2015)121;%%
  %40 citations counted in INSPIRE as of 14 Dec 2018

\bibitem{Pagura} V. P. Pagura, D. Gomez Dumm, S. Noguera, and N. Scoccola
Phys. Rev. D {\bf 95}, 034013 (2017).

\bibitem{Magdy} N. Magdy, M. Csanad and R. A. Lacey, J. Phys. G: Nucl. Part. Phys. {\bf 44}, 025101 (2017).

\bibitem{Ayala0} A. Ayala, C. A. Dominguez, S. Hernandez-Ortiz, L. A. Hernandez, M. Loewe, D. M. Paret, and R. Zamora, Phys. Rev. D {\bf 98}, 031501(R) (2018).


%\bibitem{Adams}J. Adams \textit{et al} (STAR Collaboration), Nucl. Phys. A \textbf{757}, 102 (2005).
%
%42
\bibitem{Kim} S. Chang, K. Choi, Phys. Rev. D \textbf{49}, 12 (1994).
%43
\bibitem{Braun}J. Braun, B. Klein, H. J. Pirner and A. H. Rezaeian, Phys. Rev. D \textbf{73}, 074010 (2006). 
%44

\bibitem{Abreu:2006}
  L.~M.~Abreu, M.~Gomes and A.~J.~da~Silva,
  %``Finite-size effects on the phase diagram of difermion condensates in two-dimensional four-fermion interaction models,''
  {\it  Phys. Lett. B } {\bf 642}, 551 (2006).

\bibitem{Ebert0} D. Ebert, K. G. Klimenko, A. V. Tyukov and V. Ch. Zhukovsky, {\it Phys. Rev. D} {\bf 78}, 045008 (2008).

%\cite{Abreu:2009zz}
\bibitem{Abreu:2009zz}
  L.~M.~Abreu, A.~P.~C.~Malbouisson, J.~M.~C.~Malbouisson and A.~E.~Santana,
  %``Finite-size effects on the chiral phase diagram of four-fermion models in four dimensions,''
   {\it Nucl.\ Phys.\ B } {\bf 819}, 127 (2009).
%  [arXiv:0909.5105 [hep-th]].
  %%CITATION = ARXIV:0909.5105;%%


\bibitem{Boomsma} J. K. Boomsma and D. Boer, Phys. Rev. D {\bf 80} 034019 (2009).

\bibitem{Skokov}  V. Skokov, B. Friman, E. Nakano, K. Redlich, and B.-J. Schaefer
Phys. Rev. D {\bf 82}, 034029 (2010).

\bibitem{Gatto}  R. Gatto and M. Ruggieri, Phys. Rev. D {\bf 82}, 054027 (2010). 


\bibitem{Ebert1} D. Ebert, and  K. G. Klimenko, {\it Phys. Rev. D} {\bf 82}, 025018 (2010)

\bibitem{Palhares2} L.F. Palhares, E.S. Fraga and T. Kodama, J. Phys. G \textbf{38}, 085101 (2011).
%45
%46

%\cite{Abreu:2011rj}
\bibitem{Abreu:2011rj}
  L.~M.~Abreu, A.~P.~C.~Malbouisson and J.~M.~C.~Malbouisson,
  %``Finite-size effects on the phase diagram of difermion condensates in two-dimensional four-fermion interaction models,''
  {\it  Phys.\ Rev.\ D } {\bf 83}, 025001 (2011).
 % [arXiv:1102.1860 [hep-th]].
  %%CITATION = ARXIV:1102.1860;%%

%\bibitem{Abreu2} L. M. Abreu, A. P. C. Malbouisson, J. M. C. Malbouisson and A. E. Santana, {\it Nucl. Phys. B} {\bf 819}, 127 (2011).

\bibitem{Braun3} J. Braun, B. Klein and P. Piasecki, {\it The Eur. Phys. J. C} {\bf 71}, 1576 (2011). 

\bibitem{Palhares} L. F. Palhares, E. S. Fraga and T. Kodama,
{\it J. Phys. G: Nucl. and Part. Phys.} {\bf 38}, 085101(2011). 

\bibitem{Luciano1} L. M. Abreu, A. P. C. Malbouisson and J. M. C. Malbouisson, {\it Phys. Rev. D} {\bf 84}, 065036 (2011).

\bibitem{Braun2}J. Braun, B. Klein, B.-J. Schaefer, {\it Phys. Lett. B}
{\bf 713}, 216 (2012).

\bibitem{Flachi} A. Flachi, {\it Phys. Rev. D} {\bf 86}, 104047 (2012).

\bibitem{Ebert2} D. Ebert, T. G. Khunjua, K. G. Klimenko and V. C. Zhukovsky,  {\it Int. J. Mod. Phys. A} {\bf 27}, 1250162 (2012).

\bibitem{Fraga}  E. S. Fraga and L. F.  Palhares, Phys. Rev. D {\bf 86}, 016008 (2012). 

\bibitem{Bhattacharyya1}  A. Bhattacharyya, P. Deb, S. K. Ghosh, R. Ray  and S. Sur, {\it Phys. Rev. D} {\bf 87}, 054009 (2013).


\bibitem{Abreu3} L. M. Abreu, C. A. Linhares, A. P. C. Malbouisson, J. M. C. Malbouisson,  {\it Phys. Rev. D} {\bf  88}, 107701 (2013).

\bibitem{Abreu6}L. M. Abreu, A. P. C. Malbouisson, J. M. C. Malbouisson, E. S. Nery and R. Rodrigues da Silva, Nucl. Phys. B \textbf{881}, 327-342 (2014).

\bibitem{Ebert3} D. Ebert, T. G. Khunjua, K. G. Klimenko and V. C. Zhukovsky,  {\it Phys. Rev. D} {\bf 91}, 105024 (2015). 




\bibitem{Bhattacharyya2} 
 A. Bhattacharyya, R. Ray, S. Samanta and S. Sur, {\it Phys. Rev. C} {\bf 91}, 041901 (2015)

\bibitem{Bhattacharyya3} 
 A. Bhattacharyya, R. Ray and Sur S., {\it Phys. Rev. D} {\bf 91}, 051501 (2015)

 \bibitem{Abreu4} L. M. Abreu, E. S. Nery and A. P. C. Malbouisson, Phys. Rev. D \textbf{91}, 087701 (2015).

\bibitem{Abreu5} L. M. Abreu and E. S. Nery, Int. J. Mod. Phys. A {\bf 31}, 1650128 (2016). 

\bibitem{Abreu7} L. M. Abreu, A. P. C. Malbouisson and E. S. Nery, Mod. Phys. Lett. A {\bf 31}, 1650121 (2016).



\bibitem{Bao1}S.S. Bao and H. Shen, Phys. Rev. C {\bf 93}, 025807 (2016).
%40
\bibitem{PhysRevC.96.055204} L. M. Abreu, and E. S. Nery, {\it  Phys.\ Rev.\ C }
 {\bf 96}, 055204 (2017).  

\bibitem{Samanta}S. Samanta, S. Ghosh and B. Mohanty, J. Phys. G: Nucl. Part. Phys. {\bf 45}, 075101 (2018).

\bibitem{Wu}X.H. Wu and H. Shen, Phys. Rev. C \textbf{96}, 025802 (2017).


\bibitem{Shi} C. Shi, Y. Xia, W. Jia et al., Sci. China Phys. Mech. Astron.  {\bf 61} 082021 (2018). 

\bibitem{3mats1} T. Matsubara, Prog. Theor. Phys. {\bf 14} (1955) 351.

 \bibitem{PR2014} Faqir C. Khanna, Adolfo P.C. Malbouisson, Jorge M.C.
Malbouisson and Ademir E. Santana, \textit{Quantum field theory on toroidal
topology: Algebraic structure and applications}, Phys. Rep. \textbf{539},
135-224 (2014) DOI: 10.1016/j.physrep.2014.02.002 (2014)

\bibitem{EE} E. Elizalde, {\it Ten physical applications of spectral
zeta function}, Lecture Notes in Physics (Springer-Verlag, 1995).



\bibitem{Chin} S. A. Chin, Phys. Lett. B {\bf 62}, 263 (1976).

\bibitem{Freedman1} R. A. Freedman, Phys. Lett. B {\bf 71}, 369 (1977).

\bibitem{MATSUI} T. Matsui and Brian D. Serot,  Ann. Phys. (NY) {\bf 144}, 107 (1982). 

%\cite{Ishikawa:2018yey}
\bibitem{Ishikawa:2018yey} 
  T.~Ishikawa, K.~Nakayama and K.~Suzuki,
  %``Casimir effect for nucleon parity doublets,''
  Phys.\ Rev.\ D {\bf 99}, no. 5, 054010 (2019)
  doi:10.1103/PhysRevD.99.054010
  [arXiv:1812.10964 [hep-ph]].
  %%CITATION = doi:10.1103/PhysRevD.99.054010;%%
  %1 citations counted in INSPIRE as of 02 May 2019
  
  %\cite{Buballa:1996tm}
\bibitem{Buballa:1996tm} 
  M.~Buballa,
  %``The Problem of matter stability in the Nambu-Jona-Lasinio model,''
  Nucl.\ Phys.\ A {\bf 611}, 393 (1996)
  doi:10.1016/S0375-9474(96)00314-4
  [nucl-th/9609044].
  %%CITATION = doi:10.1016/S0375-9474(96)00314-4;%%
  %109 citations counted in INSPIRE as of 02 May 2019


\bibitem{PhysRevC.83.065805} S. S. Avancini, D. P. Menezes and C. Provid\^encia, Phys. Rev. C {83}, 065805 (2011).


%\cite{Grunfeld:2014qfa}
\bibitem{Grunfeld:2014qfa} 
  A.~G.~Grunfeld, D.~P.~Menezes, M.~B.~Pinto and N.~N.~Scoccola,
  %``Phase structure of cold magnetized quark matter within the SU(3) NJL model,''
  Phys.\ Rev.\ D {\bf 90}, no. 4, 044024 (2014)
  doi:10.1103/PhysRevD.90.044024
  [arXiv:1402.4731 [hep-ph]].
  %%CITATION = doi:10.1103/PhysRevD.90.044024;%%
  %21 citations counted in INSPIRE as of 02 May 2019
%49



\end{thebibliography}
\end{document}